\documentclass[aps,prl,twocolumn,superscriptaddress,longbibliography]{revtex4-1}
\usepackage{amsmath}
\usepackage{hyperref}
\hypersetup{
	colorlinks=true,
	linkcolor=blue,
	filecolor=blue,
	citecolor = blue,      
	urlcolor=blue,
}

\usepackage{graphicx}
\usepackage{dcolumn}
\usepackage{bm}
\usepackage{natbib}
\usepackage[utf8]{inputenc}
\usepackage{units}
\usepackage{amsmath}
\usepackage{amssymb}
\usepackage{graphicx}
\usepackage{bm}
\usepackage{multirow,microtype,color,relsize,ulem}
\DeclareMathAlphabet{\mathcal}{OMS}{cmsy}{m}{n}

\usepackage[utf8]{inputenc}
\usepackage[T1]{fontenc}
\usepackage{mathptmx}
\usepackage{amsfonts}
\usepackage{amsmath}

\begin{document}

\title{Controllable interatomic interaction mediated by diffractive coupling in a cavity}

\author{Ivor Kre\v{s}i\'{c}} 
\email{ikresic@ifs.hr} 
\affiliation{Centre for Advanced Laser Techniques, Institute of Physics, Bijeni\v{c}ka cesta 46, 10000, Zagreb, Croatia}
\affiliation{Institute for Theoretical Physics, Vienna University of Technology (TU Wien), Vienna, A–1040, Austria}

\date{\today}

\begin{abstract}
Photon-mediated interaction can be used for simulating complex many-body phenomena with ultracold atoms coupled to electromagnetic modes of an optical resonator. We theoretically study a method of producing controllable interatomic interaction mediated by forward-diffracted photons circulating inside a ring cavity. One example of such a system is the three-mode cavity, where an on-axis mode can coexist with two diffracted sidebands. We demonstrate how the self-organized stripe states of a Bose-Einstein condensate (BEC) occurring in this cavity geometry can exhibit supersolid properties, due to spontaneous breaking of the Hamiltonian's continuous translational symmetry. A numerical study of the collective excitation spectrum of these states demonstrates the existence of massles and finite-gap excitations, which are identified as phase (Goldstone) and amplitude (Higgs) atomic density modes. We further demonstrate how judicious Fourier filtering of intracavity light can be used to engineer the effective atom-atom interaction profile for many cavity modes. The numerical results in this configuration show the existence of droplet array and single droplet BEC states for commensurate and incommensurate cavity modes, respectively. Diffractive coupling in a cavity is thereby introduced as a novel route towards tailoring the photon-mediated interaction of ultracold atoms. Spatial features of the self-organized optical potentials can here be tuned to scales several times larger than the pump laser wavelength, such that the corresponding atomic density distributions could be imaged and manipulated using low numerical aperture optics. These calculations and insights pave the way towards quantum simulation of exotic nonequilibrium many-body physics with condensates in a cavity.

\end{abstract}

\maketitle


Ultracold atoms coupled to modes of an optical cavity can be used for simulating complex phenomena in condensed matter physics \cite{ritsch_cold_2013,mivehvar_cavity_2021}. The effective atom-atom interaction is here mediated by cavity photons, which has the advantage of being relatively strong and highly tunable. Throughout the last two decades, various many-body effects have been explored in this context. These include the Dicke model phase transition \cite{baumann_dicke_2010}, supersolidity \cite{leonard_supersolid_2017,leonard_higgs_2017,schuster2020}, and the occurrence of phonons in a quantum gas \cite{guo_optical_2021}. 

It was recently shown that diffraction of light at a cloud of cold atoms can be used to induce self-organization in atomic density \cite{tesio_kinetic_2014,labeyrie_optomechanical_2014,robb_quantum_2015,zhang_long-range_2018,ackemann_self-organization_2021} and magnetization \cite{kresic_spontaneous_2018,labeyrie_18,labeyrie23}, in the single retroreflecting mirror configuration. This type of ordering relies on the coupling between the atoms being mediated via photons diffracted by a small angle at the atomic cloud. The work on cold and ultracold atoms extends the previous theoretical ideas regarding transverse instabilities in nonlinear media \cite{lugiato_spatial_1987,firth1990spatial,firth_transverse_1988}, where the emphasis was placed on coupling of optical waves via the material's susceptibility. 

Diffractive coupling of ultracold atoms could also be achieved using cavities, which would increase the effective strength of interactions \cite{tesio_spontaneous_2012,boquete_spontaneous_2021,ackemann_self-organization_2021}. The corresponding optical phenomena have been experimentally realized in various nonlinear media, using plano-planar cavities with intracavity lenses in the nearly self-imaging configuration, which tunes the system close to mode degeneracy \cite{esteban-martin_experimental_2004,esteban05}. Moreover, it was also shown that employing nearly confocal cavities can lead to the same behavior \cite{kreuzer90,staliunas97,ackemann2000spatial}, with the potential of entering the strong cavity-medium coupling regime.

We here describe a theoretical study of the properties of diffractive coupling in a one-dimensional Bose-Einstein condensate (BEC) coupled to electromagnetic modes of a ring cavity. We demonstrate how the profile of photon-mediated atom-atom interaction can be tailored by filtering the Fourier modes to allow propagation of light with desired transverse wavelengths. Within the mean field approximation, laser driving of the on-axis cavity mode can lead to BEC self-organization into stripe states for a setup with two cavity sidebands. We explore the supersolid properties of these stripe states arising due to spontaneous breaking of the U(1) translational symmetry of the system Hamiltonian. The Bogoliubov theory calculation reveals the presence of phase (Goldstone) and amplitude (Higgs) modes in atomic density. In the case of many cavity modes, we numerically demonstrate the existence of droplet arrays and single droplet states, for commensurate and incommensurate cavity sidebands, respectively.

Methods of controlling cavity photon-mediated interaction were recently implemented in cold atoms in a single mode cavity with an applied magnetic field gradient \cite{periwal2021programmable}, along with ultracold atoms coupled to modes of a multimode degenerate cavity \cite{vaidya_tunable-range_2018}. Another theoretically proposed method relies on using frequency combs for pumping a BEC near cavity resonances of multiple longitudinal cavity modes \cite{masalaeva23}. This Article describes a novel addition to the toolbox for control of photon-mediated interaction profiles in quantum simulations. The great advantage compared to methods of Refs. \cite{periwal2021programmable,vaidya_tunable-range_2018,masalaeva23} is that the allowed cavity mode wavevectors can be selected with considerable freedom, such that e.g. modes with incommensurate transverse wavelengths can be excited, leading to complex dynamics. 

Another advantage stems from the fact that the self-organization mechanism does not rely on collective superradiant enhancement of scattering into an optical cavity mode orthogonal to the pump laser \cite{tanji11}, and the spatial period of the patterns can be made several times larger than the pump wavelength. Experimental realization of BEC self-organization at such lengthscales would offer considerable advantages in terms of optical control and monitoring of the system dynamics as compared to the superradiant cases described in \cite{ritsch_cold_2013,mivehvar_cavity_2021}, where the spatial periods are near the diffraction limit.

\section{System Hamiltonian} 
The setup consists of a BEC placed inside a nearly degenerate ring cavity, with a laser of frequency $\omega$ driving an on-axis cavity mode with pump rate $\eta$. For strong BEC confinement along the $y$- and $z$-axes, the cloud has a cigar shape and can be approximated as one-dimensional. Moreover, for pattern lengthscales several times larger than the light's wavelength, the paraxial approximation holds and the reflection of intracavity light from the BEC can be neglected, i.e. only diffracted waves at a small angle to the pump beam direction are considered.

Using the Fourier filtering technique for photons inside the cavity \cite{jensen_manipulation_1998,labeyrie23}, one can tailor the configuration of the allowed cavity sideband modes. The cavity operator is then given by:
\begin{align}\label{eq:cavmode}
\begin{aligned}
\hat{E}(x)=\hat{a}_0+\sum_{q_j\in S} \left( \hat{a}_{j+}e^{iq_jx}+\hat{a}_{j-}e^{-iq_jx} \right),
\end{aligned}
\end{align}
where $\hat{a}_0$ is the annihilation operator of the on-axis mode and $\hat{a}_{j\pm}$ are the annihilation operators of the diffracted sideband modes with transverse wavenumbers $\pm q_j=\pm 2\pi/\Lambda_j$. The sum goes over $q_j$ values belonging to a discrete and finite set $S$.

The effective many-body Hamiltonian for the atomic and cavity degrees of freedom can be derived from the Jaynes-Cummings model, as described in Appendix A. This effective Hamiltonian has the form ($\hbar=1$):
\begin{align}\label{eq:eff_ham_main}
\begin{aligned}
\hat{H}_{eff}&=-\Delta_c\hat{a}_0^\dagger\hat{a}_0-\sum_{j}\Delta_{c,j}(\hat{a}_{j+}^\dagger\hat{a}_{j+}+\hat{a}_{j-}^\dagger\hat{a}_{j-})+i\eta(\hat{a}_0^\dagger -\hat{a}_0 )\\
+&\int\hat{\psi}^\dagger(x)\left[-\frac{1}{2m}\frac{\partial^2}{\partial x^2}+V_1(x)+U_0\hat{E}^\dagger(x)\hat{E}(x)\right]\hat{\psi}(x)dx,
\end{aligned}
\end{align}
where $\hat{\psi}(x)$ is the atomic field operator, $V_1(x)$ is the external potential, $m$ is the mass of an atom, $\Delta_c=\omega-\omega_0,\;\Delta_{c,j}=\omega-\omega_{c,j}$ are the pump detunings from the on-axis and $j$-th sideband cavity mode, respectively, while $U_0=g_0^2/\Delta_a$ is the single atom light shift, $\Delta_a=\omega-\omega_a$ is the laser-atom detuning, and $g_0$ is the atom-cavity coupling strength. The number of atoms $N$ is in the following calculations kept constant throughout the dynamical evolution.

The Hamiltonian describing the effective atom-atom interaction mediated by diffractive coupling via the cavity photons, is given by:
\begin{align}\label{eq:eff_ham_at_main}
\begin{aligned}
\hat{H}_{eff,at}&=\int\hat{\psi}^\dagger(x)\left[-\frac{1}{2m}\frac{\partial^2}{\partial x^2}+V_1(x)\right]\hat{\psi}(x)dx\\
&+\frac{1}{2}\int\hat{\psi}^\dagger(x')\hat{\psi}^\dagger(x)V_2(x,x')\hat{\psi}(x)\hat{\psi}(x')dx dx',
\end{aligned}
\end{align}
where the effective atom-atom interaction is given by:
\begin{align}\label{eq:twobodypot_main}
\begin{aligned}
V_2(x,x')&=\frac{4U_0^2\eta^2}{\bar{\Delta}_c^2+\kappa^2}\sum_{q_j\in S}\frac{\bar{\Delta}_{c,j}}{\bar{\Delta}_{c,j}^2+\kappa^2}\cos[q_j(x-x')] ,
\end{aligned}
\end{align}
and $\bar{\Delta}_c=\omega-\omega_0-NU_0$, $\bar{\Delta}_{c,j}=\omega-\omega_{c,j}-NU_0$. The effective atom-atom interaction $V_2(x,x')$ can thus be controlled by Fourier filtering to select a desired set of modes $S$.

In the following, we first explore the simple case of 2 cavity sidebands, where global all-to-all coupling between the atoms occurs. We demonstrate how such a setup can be used to simulate the phenomenon of supersolidity. We extend these studies to a multimode situation with 8 cavity sidebands, which enables us to explore the relationship between commensurability of the cavity modes and the atomic ground state solution.

\section{Supersolid properties of self-organized stripe states} 
Supersolids are an intriguing phase of matter with coexisting spatial order and superfluidity. The phenomenon has been theoretically studied during the last 50 years \cite{andreev69,chester70,leggett70,prokofev05,toner08,reppy10,boninsegni12}, however, conclusive experimental observations in the context of condensed matter physics are still lacking \cite{kim12,xiang24}. 

Several experiments have recently shown that novel insights into supersolidity can be gleaned by quantum simulation with ultracold atoms \cite{bloch_many-body_2008,mivehvar_cavity_2021,daley22}. For a  Bose-Einstein condensate (BEC), the global U(1) phase symmetry is spontaneously broken by condensation \cite{bogoliubov1947}, and spontaneous breaking of the continuous spatial (translational and/or rotational) symmetry via crystallization can lead to supersolidity. The types of interactions employed for simulating supersolidity in BECs range from cavity photon-mediated \cite{gopalakrishnan_atom-light_2010,leonard_supersolid_2017,mivehvar18,schuster2020,guo_optical_2021,masalaeva23} and strong magnetic dipolar \cite{chomaz19,botticher19,tanzi19,guo2019,norcia21,bland22,recati23} to synthetic spin-orbit \cite{li_stripe_2017}. Although such analog systems provide only simplified versions of the actual interactions present in condensed matter materials, their study can add valuable insight into the properties of the elusive supersolid phase \cite{gopalakrishnan_atom-light_2010}. 

\begin{figure}[!t]
\centering
\includegraphics[clip,width=\columnwidth]{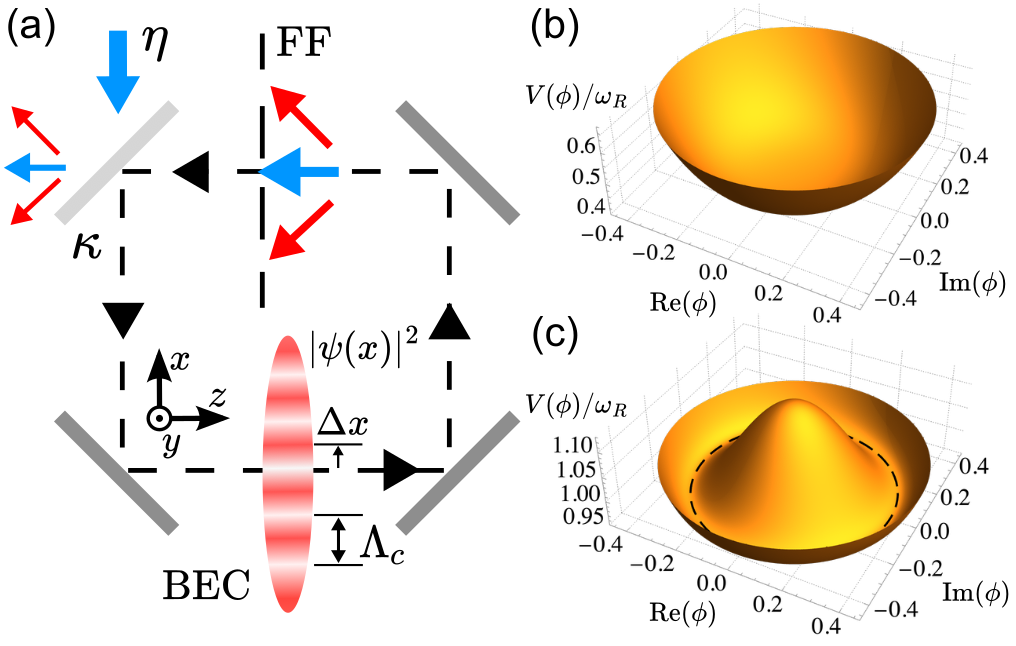}
\caption{Self-organization and supersolidity in the configuration with three cavity modes. (a) An effectively one-dimensional ultracold Bose-Einstein condensate (BEC) is placed inside a ring cavity with a linewidth $\kappa$, which is pumped by coherent on-axis light with drive amplitude $\eta$ (blue arrow). A Fourier filtering (FF) stage ensures propagation of only three cavity modes. Upon self-organization, the BEC density distribution $|\psi(x)|^2$ orders into a stripe pattern with a pattern phase displacement $\Delta x$, which can be defined as the distance of the first maximum from the origin $x=0$. The value of $\Delta x$ is selected via spontaneous symmetry breaking for a particular pattern realization, and depends on the atomic sideband phase $\mbox{arg}\phi$. Pattern lengthscale $\Lambda_c$ is tunable via Fourier filtering. Effective potential for the atoms $V(\phi)$ as a function of the complex parameter $\phi$ has (b) a circular paraboloid shape at laser pump strengths below self-organization threshold ($\eta=0.9\eta_c$), and (c) a ``sombrero" shape above threshold ($\eta=1.5\eta_c$), which is a direct consequence of the U(1) translational symmetry of the system Hamiltonian. The dashed line indicates the potential minima at $|\phi_0|=0.373$. Parameters: $(\bar{\Delta}_c,\;\bar{\Delta}_c',\; u_0 ) = (2,\;-1,\; -1)\omega_R$.}\label{Fig:1}
\end{figure}

Below we study photon-mediated supersolidity of the self-organized stripe states in the ring cavity setup, depicted in Fig. \ref{Fig:1}a). The cavity operator is now given by:
\begin{align}\label{eq:elfield_st}
\hat{E}_{st}(x)=\hat{a}_0+\hat{a}_+e^{iq_cx}+\hat{a}_-e^{-iq_cx},
\end{align}
where $\hat{a}_0$ is the on-axis mode and $\hat{a}_\pm$ the sideband mode photonic annihilation operators. The two sidebands have transverse wavenumbers $\pm q_c=\pm 2\pi/\Lambda_c$, and detunings $\Delta_c'=\omega-\omega_0'$.

The many-body Hamiltonian is here given by:
\begin{align}\label{eq:ham2}
\hat{H}_{st}=\hat{H}_{ph,st}+\int\hat{\psi}^\dagger (x)\hat{H}_{at,st}^{(1)}\hat{\psi}(x)dx,
\end{align}
where $\hat{H}_{ph,st}=-\Delta_c\hat{a}_0^\dagger \hat{a}_0-\Delta_c'(\hat{a}_+^\dagger \hat{a}_++\hat{a}_-^\dagger \hat{a}_-)+i\eta(\hat{a}_0^\dagger-\hat{a}_0)$ and the single atom Hamiltonian is given by:
\begin{align}\label{eq:hamsingleat}
\begin{aligned}
\hat{H}_{at,st}^{(1)} &=-\frac{1}{2m}\frac{\partial^2}{\partial x^2}+U_0[\hat{a}_0^\dagger \hat{a}_0 +\hat{a}_+^\dagger \hat{a}_++\hat{a}_-^\dagger \hat{a}_-+(\hat{a}_0^\dagger \hat{a}_+\\
+&\hat{a}_-^\dagger \hat{a}_0)e^{iq_cx}+(\hat{a}_0^\dagger \hat{a}_-+\hat{a}_+^\dagger \hat{a}_0)e^{-iq_cx}+\hat{a}_-^\dagger \hat{a}_+e^{2iq_cx}\\
+&\hat{a}_+^\dagger \hat{a}_- e^{-2iq_cx}].
\end{aligned}
\end{align}
Note that $\hat{H}_{st}$ is symmetric to continuous translations by distance $d$ along the $x$ axis, i.e. transformations of the form $\hat{\psi}(x)\to \hat{\psi}(x +d)$, $\hat{E}(x)\to \hat{E}(x +d)$. In the thermodynamic mean field limit, the continuous translational symmetry is spontaneously broken by the system selecting a random relative phase between the cavity sideband modes, and the phase of the periodic atomic density distribution. Note that in this case $V_2(x,x')$ varies spatially as $\cos[q_c(x-x')]$. This is analogous to the case of superradiant self-organization in a ring cavity \cite{mivehvar18}. 

Simultaneous breaking of the global U(1) phase symmetry by Bose-Einstein condensation \cite{bogoliubov1947}, and the U(1) translational symmetry by photon-mediated structuring into a density pattern means that the self-organized state of the BEC can display supersolid properties \cite{gopalakrishnan_atom-light_2010}, as studied below. For the fully quantum description, this U(1) translational symmetry of $\hat{H}_{st}$ leads to the preservation of transverse momentum of photons and atoms by the nonlinear scattering processes, which can in the transient regime lead to quantum correlations \cite{kresic2023}.

\subsection{Three-mode optomechanical model}
In the three-mode optomechanical model \cite{szirmai10}, the $\hat{\psi}(x)$ is for this setup given by:
\begin{align}\label{eq:atomfield}
\hat{\psi}(x)=\frac{1}{\sqrt{L}}\left(\hat{b}_0+\hat{b}_+e^{iq_cx}+\hat{b}_-e^{-iq_cx}\right),
\end{align}
where $\hat{b}_j$ is the bosonic annihilation operator of the $j$-th transverse atomic momentum mode and $L$ is the length of the 1D cloud. Inserting relations (\ref{eq:elfield_st}) and (\ref{eq:atomfield}) into (\ref{eq:ham2}) and performing the integration over the BEC cloud length $L$, one obtains the three-mode Hamiltonian $\hat{H}_{3M} = \hat{H}_0+\hat{H}_{FWM}$. The noninteracting part is $\hat{H}_0= -\bar{\Delta}_c\hat{a}_0^\dagger\hat{a}_0-\bar{\Delta}_c'(\hat{a}_+^\dagger\hat{a}_++\hat{a}_-^\dagger\hat{a}_-)+\omega_R(\hat{b}_+^\dagger\hat{b}_++\hat{b}_-^\dagger\hat{b}_-)+i \eta  (\hat{a}_0^\dagger- \hat{a}_0)$, where $\bar{\Delta}_c=\Delta_c-NU_0$, $\bar{\Delta}_c'=\Delta_c'-NU_0$ and $\omega_{R}=q_c^2/2m$ is the transverse photon-recoil energy, while the four-wave mixing part $\hat{H}_{FWM}$ has the form:
\begin{align}\label{eq:fwmterms}
\begin{aligned}
\hat{H}_{FWM}& =  U_0 [(\hat{a}_+^\dagger \hat{b}_-^\dagger + \hat{a}_-^\dagger \hat{b}_+^\dagger)\hat{a}_0 \hat{b}_0 \\
+&\hat{a}_0^\dagger(\hat{b}_+^\dagger \hat{a}_+ +\hat{b}_-^\dagger \hat{a}_-)\hat{b}_0 +\hat{a}_+^\dagger \hat{a}_- \hat{b}_-^\dagger
\hat{b}_+]+\mbox{H.c.}.
\end{aligned}
\end{align}
The details of the derivation of the Hamiltonian $\hat{H}_{3M}$ are given in \cite{kresic2023}.

We now derive the expression for the expectation value of the Hamiltonian, also called the effective potential $V(\phi)$ for a complex order parameter $\phi$ \cite{pekker2015,leonard_higgs_2017,guo2019}, from the microscopic Hamiltonian $\hat{H}_{3M}$ in the mean field thermodynamic limit, where $\langle \hat{O}_1 \hat{O}_2 \hat{O}_3\hat{O}_4\rangle\to\langle \hat{O}_1\rangle\langle \hat{O}_2\rangle\langle \hat{O}_3\rangle\langle \hat{O}_4\rangle$.

Writing $\langle \hat{a}_j\rangle\to \sqrt{N}\alpha_j,\:\langle \hat{b}_j\rangle\to \sqrt{N}\beta_j$ and neglecting the fourth order terms in the sidebands, which is justified near the self-organization threshold, one can calculate the energy in the photon-atom system in the mean field limit:
\begin{equation}\label{eq:meanfieldham}
\begin{aligned}
\langle \hat{H}_{3M}\rangle/N  &= -\bar{\Delta}_c |\alpha_0|^2-\bar{\Delta}_c'(|\alpha_+|^2+|\alpha_-|^2)+i\frac{\eta}{\sqrt{N}}(\alpha_0^*-\alpha_0)\\
+&\omega_R(|\beta_+|^2+|\beta_-|^2)  +u_0[\alpha_0\beta_0(\alpha_+^*\beta_-^*+\alpha_-^*\beta_+^*)\\
+&\alpha_0^*\beta_0(\alpha_+\beta_+^*+\alpha_-\beta_-^*)+\mbox{c.c.}],
\end{aligned}
\end{equation}
with $u_0=NU_0$. As $|\bar{\Delta}_c|,\:|\bar{\Delta}_c'|\gg \omega_R$, the photonic degrees of freedom adiabatically follow the atomic ones, and by setting $\frac{\partial \langle \hat{H}_{3M}\rangle}{\partial \alpha_i^*}=0$ we adiabatically eliminate them to get:
\begin{equation}\label{eq:adiabat}
\begin{aligned}
\alpha_0  = \frac{i\eta}{\sqrt{N}\bar{\Delta}_c},
\:\alpha_\pm  = \frac{u_0}{\bar{\Delta}_c'}\alpha_0\beta_0(\beta_\pm+\beta_\mp^*),
\end{aligned}
\end{equation}
where we have chosen a real-valued $\beta_0=\sqrt{1-|\beta_+|^2-|\beta_-|^2}$, while for $\alpha_0$ we have neglected the terms $\propto \frac{u_0}{\bar{\Delta}_c'}\alpha_\pm^{(*)}\beta_\pm^{(*)}$, as $|u_0|\ll|\bar{\Delta}_c'|$ and $\eta$ is near the self-organization threshold. Writing $\frac{\partial \langle \hat{H}_{3M}\rangle}{\partial \beta_\pm^*}=0$, one finds that in equilibrium $\beta_+=\beta_-^*$. 

We now introduce the complex mean field order parameter $\phi=\phi_1+i\phi_2=\beta_+=\beta_-^*$. Inserting the relations (\ref{eq:adiabat}) into the expression (\ref{eq:meanfieldham}), one finds that the effective potential $V(\phi)=\langle \hat{H}_{3M}\rangle_{eq}/N$ is rotationally symmetric, and has a Coleman-Weinberg form \cite{coleman73,kardar07,bass2021higgs} for a homogeneous (long-wavelength) order parameter $\phi$, given by:
\begin{equation}\label{eq:effective}
\begin{aligned}
V(\phi)=h_0+h_1|\phi|^2+\frac{h_2}{2}|\phi|^4,
\end{aligned}
\end{equation}
with parameters:
\begin{equation}\label{eq:constants}
\begin{aligned}
h_0=\frac{\eta^2}{N\bar{\Delta}_c},\:h_1=2\omega_R\left(1-\frac{\eta^2}{\eta_c'^2}\right),\: h_2=8\omega_R\frac{\eta^2}{\eta_c'^2}
\end{aligned}
\end{equation}
where the critical pump strength $\eta_c'$ is given by
$\eta_c'=\sqrt{-\omega_R\bar{\Delta}_c'\bar{\Delta}_c^2/(4NU_0^2)}$.
Note that $\eta_c'$ is equal to the critical pump strength $\eta_c$ \cite{kresic2023}:
\begin{equation}\label{eq:critpump}
\begin{aligned}
\eta_c=\sqrt{\frac{-\omega_R(\bar{\Delta}_c^2+\kappa^2)(\bar{\Delta}_c'^2+\kappa^2)}{4NU_0^2\bar{\Delta}_c'}},
\end{aligned}
\end{equation}
for vanishing cavity photon decay rate $\kappa$. As show in Fig. \ref{Fig:1}c,d), for $h_1>0$ ($\eta<\eta_c'$), $V(\phi)$ is a paraboloid, while for $h_1<0$ ($\eta>\eta_c'$) it deforms into a ``sombrero" shape, with minima at $|\phi|=\phi_0$, where:
\begin{equation}\label{eq:psimin}
\begin{aligned}
\phi_0=\sqrt{-\frac{h_1}{h_2}}=\frac{1}{2}\sqrt{\left(1-\frac{\eta_c'^2}{\eta^2}\right)}.
\end{aligned}
\end{equation}
The ``sombrero" shape of $V(\phi)$ above self-organization threshold is a direct consequence of the continuous translational symmetry of the Hamiltonian. After spontaneous breaking of this U(1) translational symmetry, the system Hamiltonian can support the existence of two collective excitations: one gapless related to fluctuations at constant $|\phi|$ (Goldstone or phase mode), and one with a finite energy gap related to fluctuations at a constant $\arg \phi$ (Higgs or amplitude mode) \cite{pekker2015,leonard_higgs_2017,guo2019}. The simultaneous presence of Goldstone and Higgs modes is characteristic of supersolidity, and is determined below by calculating the collective excitation spectrum using Bogoliubov-de Gennes theory.

The dynamical equations for the cavity photons and the three atomic modes can be calculated by considering the Lindblad-like evolution equations for the photon mode fields $\alpha_0(t),\alpha_\pm(t)$ and the Heisenberg equation for the three atomic motional modes $\beta_0(t),\beta_\pm(t)$, which leads to \cite{kresic2023}:
\begin{align}
\begin{aligned}
\frac{\partial \alpha_0}{\partial t} &=(i\bar{\Delta}_c-\kappa)\alpha_0-iu_0[(\beta_+^* \alpha_++\beta_-^* \alpha_-)\beta_0\\
&+\beta_0^*(\alpha_+\beta_-+\alpha_-\beta_+) ]+\frac{\eta}{\sqrt{N}},\\
\frac{\partial \alpha_\pm}{\partial t} &=(i\bar{\Delta}_c'-\kappa)\alpha_\pm-iu_0[(\beta_\mp^* \beta_0+\beta_0^* \beta_\pm)\alpha_0+\alpha_\mp\beta_\mp^* \beta_\pm],\label{eq:eqsalpha1}
\end{aligned}
\end{align}
and
\begin{align}
\begin{aligned}
\frac{\partial \beta_0}{\partial t} &=-iu_0[\alpha_0^*(\alpha_+\beta_-+\alpha_-\beta_+)+(\alpha_+^* \beta_++\alpha_-^* \beta_-)\alpha_0)],\\
\frac{\partial \beta_\pm}{\partial t} &=-i\omega_R \beta_\pm-iu_0[(\alpha_\mp^* \alpha_0+\alpha_0^* \alpha_\pm)\beta_0+\alpha_\mp^* \alpha_\pm \beta_\mp].\label{eq:eqsbeta1}
\end{aligned}
\end{align}
These equations can be used to study the dynamics of the system near the instability threshold $\eta_c$.

\begin{figure}[!t]
\centering
\includegraphics[clip,width=\columnwidth]{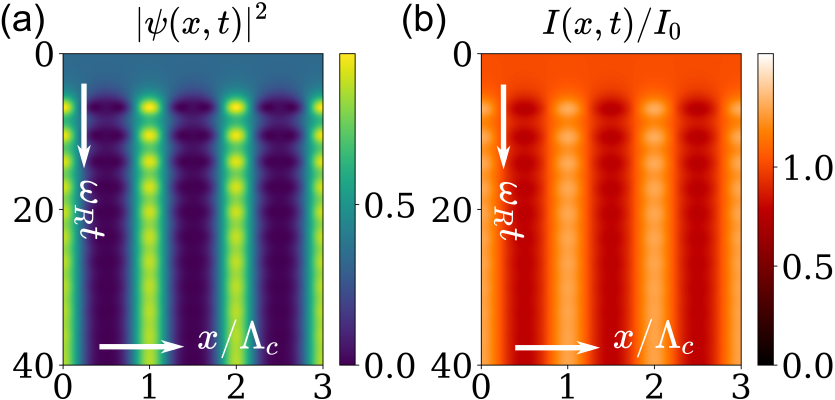}
\caption{Dynamical evolution after switching on the pump laser for (a) the atomic probability density $|\psi(x,t)|^2$ and (b) the electric field intensity profile $I(x,t)$, normalized to the steady state value of the on-axis mode intensity $I_0=|\alpha_0^0|^2$. The on-axis intracavity field initially grows on the scale $\sim 1/\kappa$ (not discernible) and approxmately reaches its steady state value $\alpha_0^0$. This homogeneous solution is not stable, and stripe patterns begin to form, with the system undergoing sloshing oscillations before reaching the steady state, where the atoms are bunched into maxima of light intensity, due to $\Delta_a<0$. The phase of the self-organized pattern was set to 0 for clarity, but can attain any value between 0 and $2\pi$ due to continuous translational symmetry of $\hat{H}_{eff}$. Simulation parameters: $\eta=1.5\eta_c$, $(\bar{\Delta}_c,\;\bar{\Delta}_c',\; u_0,\;\kappa ) = (10,\;-5,\; -1,\;5)\omega_R$.}\label{Fig:2}
\end{figure} 

In Fig. \ref{Fig:2}, we plot the temporal evolution of atomic probability distribution along the $x$-axis, given by $|\psi(x,t)|^2=|\beta_0(t)+\beta_+(t)\exp(iq_cx)+\beta_-(t)\exp(-iq_cx)|^2/L$, and the photon field profile, given by the intensity $I(x,t)=|\alpha_0(t)+\alpha_+(t)\exp(iq_cx)+\alpha_-(t)\exp(-iq_cx)|^2$, calculated by solving the Eqs. (\ref{eq:eqsalpha1},\ref{eq:eqsbeta1}). After switching on the pump laser at $t=0$, the atomic probability distribution is homogeneous, and the $\alpha_0$ quickly increases to approximately its steady state value after a time $1/\kappa$. This homogeneous solution soon becomes unstable, and the stripe pattern begins to form. At the onset of the instability, the system undergoes sloshing motion, marked by continuous oscillation between the bunched and homogeneous atomic and electric field pattern, studied e.g. in \cite{tesio_self-organization_2014}. After the oscillations decay, a striped steady state is reached, with atoms bunching into the peaks of the electric field intensity due to the drive being red-detuned from the optical transition of the atoms, i.e. $U_0<0$. 

\subsection{Gross-Pitaevskii-type equation}
Following the approach of \cite{nagy08,mivehvar18}, we here move away from the three-mode model, which leads to equations of the Gross-Pitaevskii type for the atomic wavefunction. Using the same approximations as above with the Hamiltonian $\hat{H}_{st}$, we get the dynamical equations for the photon mode fields $\alpha_0(t),\alpha_\pm(t)$ and atomic wavefunction $\psi(x,t)$:
\begin{align}\label{eq:eqsallfull}
\begin{aligned}
\frac{\partial \alpha_0}{\partial t} &=(i\bar{\Delta}_c-\kappa) \alpha_0 -iu_0[\alpha_+\mathcal{I}_++\alpha_-\mathcal{I}_-]+\frac{\eta}{\sqrt{N}},\\
\frac{\partial \alpha_\pm}{\partial t} &=(i\bar{\Delta}_c'-\kappa) \alpha_\pm -iu_0[\alpha_0\mathcal{I}_\mp+\alpha_\mp \mathcal{I}_{2\mp}],\\
 i\frac{\partial \psi}{\partial t} &=-\frac{1}{2m}\frac{\partial^2 \psi}{\partial x^2} +u_0\left[|\alpha_0|^2+|\alpha_+|^2+|\alpha_-|^2\right.\\
 +&  (\alpha_0^*\alpha_-+\alpha_+^*\alpha_0)e^{-iq_cx}+\alpha_+^*\alpha_-e^{-2iq_cx}\\
 +&\left.(\alpha_0^*\alpha_++\alpha_-^*\alpha_0)e^{iq_cx}+\alpha_-^*\alpha_+e^{2iq_cx}\right]\psi,
 \end{aligned}
\end{align}
where the integrals $\;\mathcal{I}_\pm=\int e^{\pm iq_c x}|\psi(x,t)|^2dx$, $\mathcal{I}_{2\pm}=\int e^{\pm 2iq_c x}|\psi(x,t)|^2dx$ quantify the ordering of atoms into the optical potential minima.


We find the stationary ground state solution $e^{-i\mu_0 t}\psi_0(x)$ of Eqs. (\ref{eq:eqsallfull}) by self-consistently solving the equations via setting $\partial_t\alpha_j=0$ for $j=0,\pm$ and $i\partial_t\psi=\mu\psi$ \cite{masalaeva23}. Neglecting the BEC medium boundary effects, the $\Lambda_c$ periodicity of the optical potential arising from the cavity photon modes in Eq. (\ref{eq:elfield_st}) allows us to solve the equations on a box of length $\Lambda_c$, applying periodic boundary conditions. 

\begin{figure}[!t]
\centering
\includegraphics[clip,width=\columnwidth]{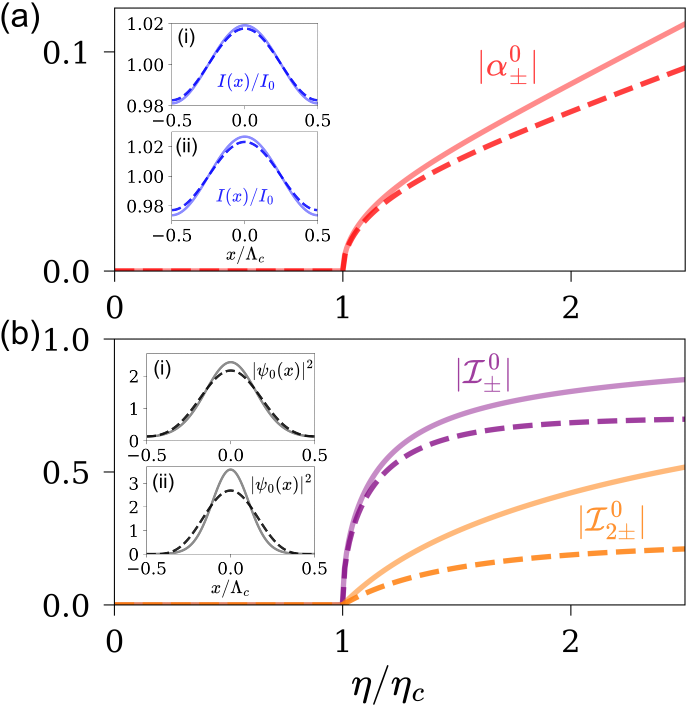}
\caption{Steady state values of (a) $|\alpha_\pm^0|$ and (b) $|\mathcal{I}_\pm^0|$, $|\mathcal{I}_{2\pm}^0|$ versus $\eta$. The insets depict the (a) intracavity electric field intensity $I(x)/I_0$ and (b) probability distribution $|\psi_0(x)|^2$, where for (i) $\eta=1.2\eta_c$ and for (ii) $\eta=1.8\eta_c$, with the phase of the self-organized pattern set to 0 for clarity. The solid lines in all subplots represent the results for the steady state solutions of the Eqs. (\ref{eq:eqsallfull}), while the dashed lines are for the solutions of the three-mode model, where $\mathcal{I}_\pm^0=\beta_\mp^0(\beta_0^{0})^*+\beta_0^0(\beta_\pm^{0})^*$, and $\mathcal{I}_{2\pm}^0=\beta_\mp^0(\beta_\pm^0)^*$. Simulation parameters: $(\bar{\Delta}_c,\;\bar{\Delta}_c',\; u_0,\;\kappa ) = (119,\;-89,\; -1,\;50)\omega_R$.}\label{Fig:3}
\end{figure}

In Fig. \ref{Fig:3}, we compare the steady-state results for the three mode model and the full system of Eqs. (\ref{eq:eqsallfull}). Both the sideband fields $\alpha_\pm^0$ and the integrals $\mathcal{I}_\pm^0,\mathcal{I}_{2\pm}^0$ vanish below threshold. For pump rates $\eta\gtrsim\eta_c$, the quantities acquire finite values, which are approximately equal for the three-mode model and the Eqs. (\ref{eq:eqsallfull}). For these $\eta$ values, both the electric field intensity distributions and the probability distributions are also approximately equal for the two models. The three-mode model gives thus a very good approximation of the system behavior in this regime. 

Increasing $\eta$ higher above threshold, the modulation depth of the self-organized cavity photon standing wave becomes quadratically larger, which can be seen from the inset of Fig. \ref{Fig:3}a) by noting that $I_0\approx \eta^2/N/(\bar{\Delta}_c^2+\kappa^2)$. The atoms are thus more strongly localized in the optical potential wells, and their probability distribution $|\psi_0(x)|^2$ becomes noticeably different than the prediction for the three-mode approximation. This is also reflected in the increase of discrepancy of the $|\alpha_\pm^0|$, $|\mathcal{I}_\pm^0|$ and $|\mathcal{I}_{2\pm}^0|$ values for the two models. Including higher order atomic momenta thus becomes necessary for quantitatively describing the steady state system behavior for large $\eta/\eta_c$ values.


\subsection{Collective excitation spectrum} 
We here use the Bogoliubov-de Gennes approach outlined in \cite{nagy08,szirmai10,masalaeva23} to calculate the collective excitation spectrum. We consider solutions of the form $\psi(x,t)=e^{-i\mu_0 t}[\psi_0(x)+\delta\psi(x,t)]$, $\alpha_j(t)=\alpha_j^0+\delta\alpha_j(t)$, where $\delta\psi(x,t)=\delta\psi^+(x)e^{-i\Omega t}+[\delta\psi^{-}(x)]^*e^{i\Omega^* t}$ and $\delta\alpha_j(t)=\delta\alpha_j^+e^{-i\Omega t}+[\delta\alpha_j^{-}]^*e^{i\Omega^* t}$, for $j=0,\pm$. Inserting these relations into Eq. (\ref{eq:eqsallfull}) and collecting the $e^{-i\Omega t}$, $e^{i\Omega^* t}$ terms, the collective excitation spectrum is now determined by the eigenvalue equation:
\begin{align}\label{eq:atomperturbs}
\hat{\mbox{M}}\mathbf{R} &=\Omega\mathbf{R},
\end{align}
with the non-Hermitian excitation matrix $\hat{\mbox{M}}$ given by Eq. (\ref{eq:eqsallperturb_eigenfinal}), and the excitation vector defined as: 
\begin{equation}\label{eq:perturbvector}
\mathbf{R}=[\delta\alpha_0^+,\delta\alpha_0^-,\delta\alpha_+^+,\delta\alpha_+^-,\delta\alpha_-^+,\delta\alpha_-^-,\delta\psi^+(x),\delta\psi^-(x)]^{\top}.
\end{equation}
We numerically calculate the spectrum of $\hat{\mbox{M}}$ by using a spatial grid of length $\Lambda_c$ with 256 points and periodic boundary conditions. Due to the symmetry properties of $\hat{\mbox{M}}$, its eigenvalues appear in pairs with equal imaginary parts and real parts with opposite signs \cite{nagy08,szirmai10}, and we here plot only the part of the spectrum with positive real parts.

\begin{figure}[!t]
\centering
\includegraphics[clip,width=\columnwidth]{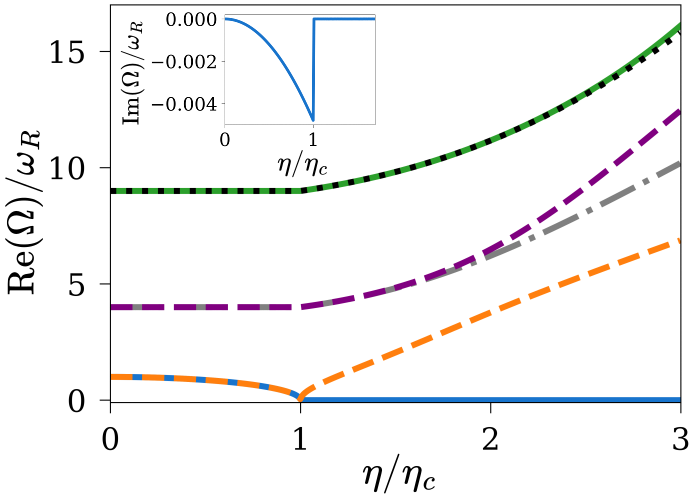}
\caption{Real parts of the lowest 6 eigenvalues of the collective excitation matrix $\hat{\mbox{M}}$ vs. pump rate $\eta$. The plotted modes represent condensate density excitations at frequencies $\omega_R,4\omega_R,9\omega_R$ for $\eta\ll\eta_c$, and the lowest energy photon modes appear at $\mbox{Re}(\Omega)=-\bar{\Delta}_c'=89\omega_R$. Above threshold, one of the two lowest modes becomes gappless (blue, solid), indicating Goldstone mode behavior, while the other mode obtains an energy gap increasing with $\eta$ (orange, dashed), indicating Higgs mode behavior, as a direct consequence of spontaneous breaking of the U(1) translational symmetry of the Hamiltonian $\hat{H}_{eff}$ by self-organization \cite{leonard_higgs_2017,guo2019}. Inset: The decay rate $-\mbox{Im}(\Omega)$ of the Goldstone mode vanishes above threshold, as a consequence of the fact that the continuous translational symmetry of the system equations is not broken by cavity photon decay at rate $\kappa$. Simulation parameters are given in the caption of Fig. \ref{Fig:3}.}\label{Fig:4}
\end{figure} 

In Fig. \ref{Fig:4}, we plot the first 6 eigenvalues of $\hat{\mbox{M}}$ for scanning $\eta$, excluding the zero frequency mode which gives rise to phase diffusion of the homogeneous part of the condensate for all $\eta$ \cite{nagy_critical_2011}. At $\eta\ll\eta_c$, the collective condensate mode pairs have frequencies $\omega_R,4\omega_R,9\omega_R$, which corresponds to the spectrum of a condensate in a box \cite{nagy08}. The lowest energy cavity photon  modes have for the used parameters the frequency $\sim -\bar{\Delta}_c'=89\omega_R$, which is not within the range of the plot. 

Near self-organization threshold pump rate $\eta_c$, the lowest two mode frequencies approach zero, and moving beyond $\eta_c$ leads to the splitting of the modes into a gapless mode, indicating a phase (Goldstone) mode, and a mode with a finite energy gap. The frequency of the finite gap mode vanishes for $\eta\to\eta_c$, indicating an amplitude (Higgs) mode excitation with fluctuations in the strength of the atomic density modulation \cite{pekker2015}. The simultaneous presence of Goldstone and Higgs modes in the collective excitation spectrum is a signature of BEC supersolidity \cite{leonard_higgs_2017,guo2019}. In contrast to the model of \cite{leonard_higgs_2017,lang17}, the continuous translational symmetry is here exact, such that no finite gap appears for the Goldstone mode. 

Note that the cavity photon decay with rate $\kappa$ may cause finite values of the decay rates $-\mbox{Im}(\Omega)$ of the collective excitations. However, for the Goldstone mode the decay rate is finite only for $\eta<\eta_c$, and the damping vanishes above threshold (see inset of Fig. \ref{Fig:4}). This can be explained by the fact that cavity photon decay does not break the U(1) translational symmetry of the Lindblad evolution equations for the photon modes \cite{kresic2023}, such that finite $\kappa$ will not dampen the Goldstone mode appearing after spontaneous symmetry breaking of U(1) translational symmetry in the thermodynamic limit. The same feature is also present in the model studied in \cite{mivehvar18}. 

\section{Commensurate and incommensurate modes in a multimode cavity} 
The great majority of work on quantum simulation with atoms coupled to optical resonators has involved coupling to a single or two cavity modes, which gives rise to global interaction between the atoms \cite{mivehvar_cavity_2021}. However, many of the Hamiltonians of interest to the condensed matter physics community, such as models of the glassy phase \cite{strack11,angelone16,marsh24} or frustrated magnetism \cite{hung16}, rely on more exotic types of interaction.

We here study a simple example where the effective atom-atom interaction has a more complex spatial profile compared to the global interaction case described in the previous section. This modification is achieved by tailoring the distribution of multiple sideband modes circulating inside the cavity. The Gross-Pitaevskii type equation for the atoms is now given by
\begin{align}\label{eq:gpemultimode}
\begin{aligned}
 i\frac{\partial \psi}{\partial t} &=-\frac{1}{2m}\frac{\partial^2 \psi}{\partial x^2}+V_1(x)\psi +u_0\{|\alpha_0|^2+\sum_{q_j\in S}[|\alpha_{j+}|^2+|\alpha_{j-}|^2\\
 +&  (\alpha_0^*\alpha_{j-}+\alpha_{j+}^*\alpha_0)e^{-iq_jx}+\alpha_{j+}^*\alpha_{j-}e^{-2iq_jx}\\
 +&(\alpha_0^*\alpha_{j+}+\alpha_{j-}^*\alpha_0)e^{iq_jx}+\alpha_{j-}^*\alpha_{j+}e^{2iq_jx}]\}\psi,
\end{aligned}
\end{align}
where $V_1(x)$ is an external box-shaped potential, which leads to hard wall boundary conditions for $\psi(x)$ at the edges of the system. As the on-axis mode is driven by a pump laser, we here approximate it by $\alpha_0\simeq i\eta/[\sqrt{N}(\bar{\Delta}_c+i\kappa)]$, see Appendix B. For large $\kappa$ and/or $|\bar{\Delta}_{c,j}|$, and $|\Delta_a|$ values, sideband modes can then be adiabatically eliminated and approximated by
\begin{align}\label{eq:aj_op}
\begin{aligned}
\alpha_{j\pm}&\simeq\frac{U_0\alpha_0}{\bar{\Delta}_{c,j}+i\kappa}\int e^{\mp iq_jx}|\psi(x)|^2dx,
\end{aligned}
\end{align}
as also shown in the Appendix B. 

\begin{figure}[!t]
\centering
\includegraphics[clip,width=\columnwidth]{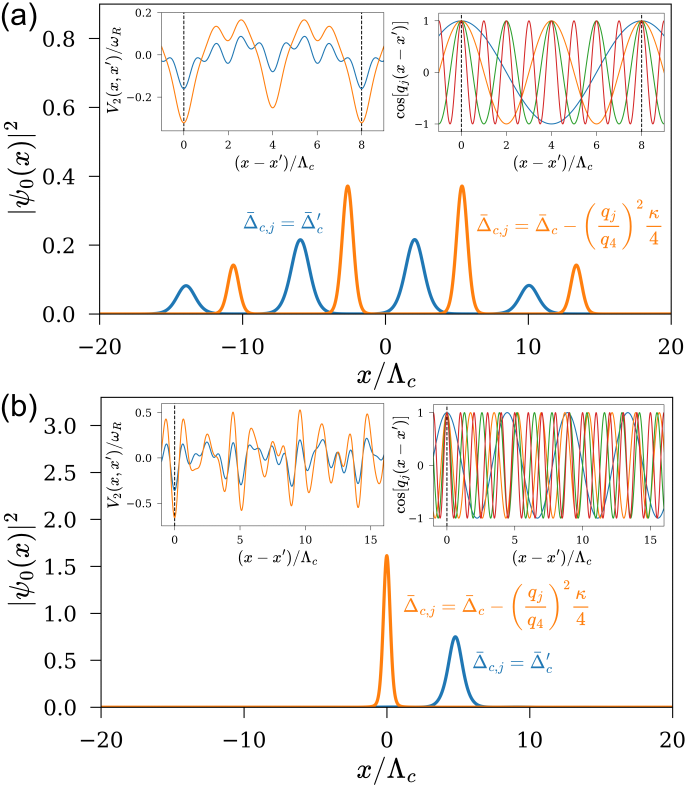}
\caption{Results for a multimode cavity with 9 cavity modes. (a) $|\psi_0|^2$ for commensurate cavity modes with $q_j=(1,0.5,0.25,0.125)q_c$ (blue, constant $\bar{\Delta}_{c,j}$ and orange, variable $\bar{\Delta}_{c,j}$). (b) $|\psi_0|^2$ for incommensurate cavity modes with $q_j=(1,0.775,0.575,0.225)q_c$ (blue, constant $\bar{\Delta}_{c,j}$ and orange, variable $\bar{\Delta}_{c,j}$). Left inset: atom-atom interaction profile $V_2(x,x')$ with $U_0\to u_0$ (blue, constant $\bar{\Delta}_{c,j}$ and orange, variable $\bar{\Delta}_{c,j}$). Right inset: cosine functions for modes with different $q_j$ values (red, $q_1$, green, $q_2$, orange, $q_3$, blue, $q_4$). Simulation parameters: $(\bar{\Delta}_c,\;\bar{\Delta}_c',\; u_0,\;\kappa ) = (3,\;-2,\; -0.1,\;20)\omega_R$, (a) $\eta=290\omega_R$ (blue), $\eta=270\omega_R$ (orange), (b) $\eta=430\omega_R$ (blue), $\eta=350\omega_R$ (orange). For constant $\bar{\Delta}_{c,j}$, the value is $\bar{\Delta}_{c,j}=\bar{\Delta}_c'$. For variable $\bar{\Delta}_{c,j}$ the value is changed as $\bar{\Delta}_{c,j}=\bar{\Delta}_c-(q_j/q_4)^2\kappa/4$. Hard wall boundary conditions were imposed at the system ends at $x=\pm 20$.}\label{Fig:5}
\end{figure}

For the configurations studied here, 9 cavity modes are selected by Fourier filtering. In the first example, the wavevenumbers are commensurate and given by $q_j=(1,0.5,0.25,0.125)q_c$. The probability distribution for a realization of the ground state wavefunction $\psi_0(x)$ is shown in Fig. \ref{Fig:5}a). 
For these commensurate transverse wavelengths, the sideband cavity modes constructively interfere at lengthscales of $8\Lambda_c$, which is also the periodicity of the minima of the atom-atom interaction profile $V_2(x,x')$. The atoms can thus order into a droplet array of periodicity $8\Lambda_c$, which simultaneously increases the diffraction into all of the allowed cavity modes and minimizes the atomic potential energy.

In contrast, when the modes are incommensurate, such ordering in $\psi_0(x)$ does not occur, as shown in the example in Fig. \ref{Fig:5}b), where $q_j=(1,0.775,0.575,0.225)q_c$. Here, a single constructive interference peak between all of the allowed modes occurs for $|x-x'|=0$. It is thus not possible to simultaneously enhance the scattering into all cavity modes for atoms arranged in a periodic array. The ground state solution is now a single droplet state.

Detunings $\Delta_{c,j}$ depend on the diffractive phase shifts of the sideband modes propagating in the system \cite{ackemann_self-organization_2021,kresic2023}. In Fig. \ref{Fig:5}, we plot the case for which $\bar{\Delta}_{c,j}$ is equal for all sidebands, and the case for which $\bar{\Delta}_{c,j}$ depends on $q_j$. For the latter case, near cavity degeneracy the quadratic relationship $\omega_{c,j}-\omega_0\propto q_j^2$ holds \cite{esteban-martin_experimental_2004,kresic2023}, which is here used to determine the detunings $\bar{\Delta}_{c,j}$. The two cases have the same ground state solutions. The depth of the $V_2(x,x')$ minima is larger for the parameters of the case with variable $\bar{\Delta}_{c,j}$, such that the localization of the atoms at the optical potential minima is here stronger. Using wavefront shaping elements inside the cavity feedback loop, the cavity sideband phase shifts can be controlled in order to tailor the distribution of $\bar{\Delta}_{c,j}$ values and, in combination with selecting the $q_j$ distribution, engineer the shape of $V_2(x,x')$, which is an intriguing prospect for applications in quantum simulation.

\section{Conclusions}
A novel method for providing tailored photon-mediated atom-atom interaction in a BEC was studied theoretically and numerically. The method is based on diffractive coupling by electromagnetic modes of a ring cavity. In the first example of self-organization into stripe states, it is shown how the continuous translational symmetry of the system Hamiltonian is manifested in the ``sombrero" shape of the effective mean field potential for the atomic degrees of freedom above self-organization threshold. This symmetry is spontanously broken in the thermodynamic limit, by the system selecting a particular phase for the steady state sinusoidal pattern in atomic density and intracavity light field profile. Supersolid properties are exhibited by the simultaneous presence of Goldstone and Higgs modes in the numerically calculated collective excitation spectrum. The calculations corroborate previous remarks regarding supersolidity of optomechanical patterns in a BEC, observed for simulations in the single retroreflecting mirror configuration \cite{robb_quantum_2015}, which is a setup similar to the ring cavity scheme \cite{ackemann_self-organization_2021}. It should be readily possible to extend the problem into two dimensional geometries, where an additional rotational U(1) symmetry will be broken by self-organization, which relates to the recent interest in supersolidity of dipolar condensates in two dimensions \cite{norcia21,bland22}.

For highly elongated clouds \cite{nguyen2017formation,lim21}, spatial features can be tuned to values of tens of microns, which would enable real space monitoring of the condensate dynamics. This also opens the possibility of modifying the photon-mediated interatomic interaction by placing optical elements with spatially tailored profiles within the optical feedback loop of the cavity.

Going beyond global all-to-all atom-atom coupling occurring for three cavity modes, we also studied the case of a multimode cavity with a tailored distribution of photonic sidebands. For commensurate transverse wavenumbers $q_j$, the ground state is a droplet array, which simultaneously enhances scattering into all of the allowed cavity modes. In contrast, for incommensurate modes only one atom-atom potential minimum occurs, at $|x-x'|=0$. The resulting ground state solution is a single droplet state. These example situations demonstrate the versatility of the proposed experimental setup. A wide variety of interaction profiles $V_2(x,x')$ can be implemented in this context, which may lead to non-equilibrium realizations of models of spin glasses \cite{strack11,angelone16,marsh24} or frustrated magnetism \cite{hung16}.

\textit{Acknowledgements.} I thank Natalia Masalaeva for sharing her code and for many helpful discussions, and Thorsten Ackemann for comments on the early version of the manuscript. This work was funded by the Austrian Science Fund (FWF Meitner-Programm, Project No. M3011) and the Austrian Academy of Sciences (ESQ Discovery Programm, Project: “Using self-organization of ultracold atoms in emerging quantum technologies”). I acknowledge the support of the project KODYN financed by the European Union through the National Recovery and Resilience Plan 2021-2026 (NRPP). In addition, this work was supported by the project Centre for Advanced Laser Techniques (CALT), co-funded by the European Union through the European Regional Development Fund under the Competitiveness and Cohesion Operational Programme (Grant No. KK.01.1.1.05.0001). The computational results presented here have been achieved using the Vienna Scientific Cluster (VSC). The dynamical evolution equations were solved numerically by using the open-source framework QuantumOptics.jl \cite{kramer_quantumopticsjl_2018}.

\section{Appendix A: Derivation of the effective Hamiltonian}
\label{appendix:A}
\setcounter{equation}{0}
\renewcommand{\theequation}{A\arabic{equation}}
\renewcommand{\theHsection}{A\arabic{section}}

The effective Hamiltonian (\ref{eq:eff_ham_main}) can be derived from the Jaynes-Cummings model, similarly to the procedure given in Refs. \cite{maschler_ultracold_2008,masalaeva23}. We describe the atom-light interaction within the domain of paraxial approximation, where only forward-directed scattering of light by the condensate can occur, such that only counterclockwise- or clockwise-propagating modes can be excited by the instability. We can thus concentrate on a one-dimensional model, where the cavity mode operator at the BEC medium is given by: 
\begin{align}\label{eq:cavmode2}
\begin{aligned}
\hat{E}(x)=\hat{a}_0+\sum_{q_j\in S} \left( \hat{a}_{j+}e^{iq_jx}+\hat{a}_{j-}e^{-iq_jx} \right),
\end{aligned}
\end{align}
where $\hat{a}_0$ is the annihilation operator of the on-axis plane wave mode, and $\hat{a}_{j\pm}$ are the annihilation operators of the diffracted sidebands with transverse wavenumbers $\pm q_j=\pm 2\pi/\Lambda_j$. The allowed sideband wavenumbers $q_j$ are part of a bounded discrete set $S$, and the values contained within $S$ can be experimentally tailored via Fourier filtering \cite{jensen_manipulation_1998,labeyrie23}.

In the dipole and rotating wave approximations, the single atom Jaynes-Cummings Hamiltonian of the system is given by ($\hbar=1$):
\begin{widetext}
\begin{align}\label{eq:h1at_1}
\begin{aligned}
\hat{\mathcal{H}}^{(1)} &=-\frac{1}{2m}\frac{\partial^2}{\partial x^2}+ \omega_a\hat{\sigma}_{ee}+\{[g_0\hat{a}_0+\sum_{q_j\in S}g_j( \hat{a}_{j+}e^{iq_jx}+\hat{a}_{j-}e^{-iq_jx})]\hat{\sigma}_{eg}+\mbox{H.c.}\}\\
+& V_e(x)\hat{\sigma}_{ee}+ V_g(x)\hat{\sigma}_{gg}+\omega_0\hat{a}_0^\dagger\hat{a}_0+\sum_{j}\omega_j(\hat{a}_{j+}^\dagger\hat{a}_{j+}+\hat{a}_{j-}^\dagger\hat{a}_{j-})+i\eta(\hat{a}_0^\dagger e^{-i\omega t}-\hat{a}_0 e^{i\omega t}),
\end{aligned}
\end{align}
\end{widetext}
where $m$ is the mass of the atom, $\omega_a$, $\omega$, $\omega_j$ are the frequencies of the atomic two-level transition, drive laser and the $j$-th cavity mode, respectively, $g_j$ is the atom-cavity coupling strength of the $j$-th mode, $\hat{\sigma}_{kk'}=|k\rangle\langle k'|$ is the atomic optical transition operator, $V_{g,e}(x)$ are the external trapping potentials for atoms in the ground and excited states, respectively, $\eta\in\mathbb{R}$ is the laser pump rate and the sum over $j$ implies Fourier filtering of $q_j$ was performed. In the following we approximate all $g_j$ values by $g_0\in\mathbb{R}$. Note that we here approximate the direct atom-atom scattering in the BEC to be sufficiently small to be neglected, i.e. the phenomena studied in this Article are based only on photon-mediated interaction between the atoms.

The Hamiltonian (\ref{eq:h1at_1}) can be converted into a time-independent form
\begin{widetext}
\begin{align}\label{eq:h1at_2}
\begin{aligned}
\hat{\tilde{\mathcal{H}}}^{(1)} &=\hat{U}\hat{\mathcal{H}}^{(1)}\hat{U}^\dagger+i(\partial_t\hat{U})\hat{U}^\dagger\\
&=-\frac{1}{2m}\frac{\partial^2}{\partial x^2}-\Delta_a\hat{\sigma}_{ee}+(g_0\hat{E}(x)\hat{\sigma}_{eg}+\mbox{H.c.})+V_e(x)\hat{\sigma}_{ee}+ V_g(x)\hat{\sigma}_{gg}-\Delta_c\hat{a}_0^\dagger\hat{a}_0-\sum_{j}\Delta_{c,j}(\hat{a}_{j+}^\dagger\hat{a}_{j+}+\hat{a}_{j-}^\dagger\hat{a}_{j-})+i\eta(\hat{a}_0^\dagger -\hat{a}_0 ),
\end{aligned}
\end{align}
\end{widetext}
where $\Delta_a=\omega-\omega_a$, $\Delta_c=\omega-\omega_0$, $\Delta_{c,j}=\omega-\omega_j$, by using
\begin{align}\label{eq:h1at_2}
\begin{aligned}
\hat{U} &=\exp\left\{i\omega t\left[\hat{\sigma}_{ee}+\hat{a}_0^\dagger\hat{a}_0+\sum_{j}(\hat{a}_{j+}^\dagger\hat{a}_{j+}+\hat{a}_{j-}^\dagger\hat{a}_{j-})\right]\right\}.
\end{aligned}
\end{align}
The corresponding many-body Hamiltonian of the system is given by:
\begin{widetext}
\begin{align}\label{eq:ham_manybody_1}
\begin{aligned}
\hat{H} &=\int \hat{\Psi}_g^\dagger(x)\left[-\frac{1}{2m}\frac{\partial^2}{\partial x^2}+V_g(x)\right]\hat{\Psi}_g(x) dx+\int \hat{\Psi}_e^\dagger(x)\left[-\frac{1}{2m}\frac{\partial^2}{\partial x^2}+V_e(x)-\Delta_a\right]\hat{\Psi}_e(x) dx\\
+&\int\left( g_0\hat{\Psi}_e^\dagger(x)\hat{E}(x)\hat{\Psi}_g(x) +\mbox{H.c.}\right)dx -\Delta_c\hat{a}_0^\dagger\hat{a}_0-\sum_{j}\Delta_{c,j}(\hat{a}_{j+}^\dagger\hat{a}_{j+}+\hat{a}_{j-}^\dagger\hat{a}_{j-})+i\eta(\hat{a}_0^\dagger -\hat{a}_0 ),
\end{aligned}
\end{align}
\end{widetext}
where
\begin{align}\label{eq:commutrel_at}
\begin{aligned}
&[\hat{\Psi}_{k}(x),\hat{\Psi}_{k'}^\dagger(x')]=\delta(x-x')\delta_{k,k'},\\
&[\hat{\Psi}_{k}(x),\hat{\Psi}_{k'}(x')]=[\hat{\Psi}_{k}^\dagger(x),\hat{\Psi}_{k'}^\dagger(x')]=0.
\end{aligned}
\end{align}
The Heisenberg equation for $\hat{\Psi}_e(x)$ is given by 
\begin{align}\label{eq:eqpsie}
\begin{aligned}
i\frac{\partial\hat{\Psi}_e(x)}{\partial t}=\left[-\frac{1}{2m}\frac{\partial^2}{\partial x^2}+V_e(x)-\Delta_a\right]\hat{\Psi}_e(x)+g_0\hat{E}(x)\hat{\Psi}_g(x)
\end{aligned}
\end{align}
In the limit of large $\Delta_a$, the excited state dynamics reaches steady state much more quickly than the ground state. The kinetic energy and external potential terms can be neglected in comparison to $\Delta_a$, and $\hat{\Psi}_e(x)$ can be adiabatically eliminated to give:
\begin{align}\label{eq:eqpsiesteady}
\begin{aligned}
\hat{\Psi}_e(x)\simeq\frac{1}{\Delta_a}g_0\hat{E}(x)\hat{\Psi}_g(x).
\end{aligned}
\end{align}
The Heisenberg equations for the cavity operators $\hat{a}_0,\hat{a}_j$ (including a phenomenological cavity decay constant $\kappa$) and the atomic field $\hat{\Psi}_g(x)\equiv\hat{\psi}(x)$ can be derived from (\ref{eq:ham_manybody_1}), which after inserting (\ref{eq:eqpsiesteady}) leads to:
\begin{align}\label{eq:heiseqs1}
\begin{aligned}
\frac{\partial \hat{a}_0}{\partial t}&=(i\Delta_c-\kappa)\hat{a}_0
-iU_0\int\hat{\psi}^\dagger(x)\hat{E}(x)\hat{\psi}(x)dx+\eta,\\
\frac{\partial \hat{a}_{j\pm}}{\partial t}&=(i\Delta_{c,j}-\kappa)\hat{a}_{j\pm}
-iU_0\int\hat{\psi}^\dagger(x)\hat{E}(x)e^{\mp iq_jx}\hat{\psi}(x)dx,\\
\frac{\partial \hat{\psi}(x)}{\partial t}&=\left[-\frac{1}{2m}\frac{\partial^2}{\partial x^2}+V_1(x)+U_0\hat{E}^\dagger(x)\hat{E}(x)\right]\hat{\psi}(x).
\end{aligned}
\end{align}
Eqs. (\ref{eq:heiseqs1}) can be derived from the Heisenberg equation with the effective Hamiltonian $\hat{H}_{eff}$, given by:
\begin{align}\label{eq:eff_ham2}
\begin{aligned}
\hat{H}_{eff}&=-\Delta_c\hat{a}_0^\dagger\hat{a}_0-\sum_{j}\Delta_{c,j}(\hat{a}_{j+}^\dagger\hat{a}_{j+}+\hat{a}_{j-}^\dagger\hat{a}_{j-})+i\eta(\hat{a}_0^\dagger -\hat{a}_0 )\\
+&\int\hat{\psi}^\dagger(x)\left[-\frac{1}{2m}\frac{\partial^2}{\partial x^2}+V_1(x)+U_0\hat{E}^\dagger(x)\hat{E}(x)\right]\hat{\psi}(x)dx.
\end{aligned}
\end{align}

\section{Appendix B: Photon-mediated interatomic interaction}
In the limit of large $|\Delta_{c,j}|$ and/or $\kappa$, the cavity photon operators can be adiabatically eliminated. The system can then be described by a Hamiltonian of the form:
\begin{align}\label{eq:eff_ham3_2}
\begin{aligned}
\hat{H}_{eff,at}&=\int\hat{\psi}^\dagger(x)\left[-\frac{1}{2m}\frac{\partial^2}{\partial x^2}+V_1(x)\right]\hat{\psi}(x)dx\\
&+\frac{1}{2}\int\hat{\psi}^\dagger(x')\hat{\psi}^\dagger(x)V_2(x,x')\hat{\psi}(x)\hat{\psi}(x')dx dx',
\end{aligned}
\end{align}
where $V_2(x,x')$ is the effective atom-atom interaction mediated by cavity photons. 

To determine the form of $V_2(x,x')$, we first adiabatically eliminate the on-axis mode, which leads to: 
\begin{align}\label{eq:ao_op}
\begin{aligned}
\hat{a}_0&\simeq\frac{i\eta}{\bar{\Delta}_c+i\kappa}+\mathcal{O}\left(\frac{1}{\Delta_a}\right),
\end{aligned}
\end{align}
where $\bar{\Delta}_c=\omega-\omega_0-NU_0$, and we neglect the terms $\mathcal{O}(1/\Delta_a)$. As the on-axis mode is driven by a laser, one can approximate $\hat{a}_0$ by $\langle\hat{a}_0\rangle\to \sqrt{N}\alpha_0$, where $\alpha_0\simeq i\eta/[\sqrt{N}(\bar{\Delta}_c+i\kappa)]$ \cite{kresic2023}. The sideband cavity photon operators can then be adiabatically eliminated to give: 
\begin{align}\label{eq:aj_op}
\begin{aligned}
\hat{a}_{j\pm}&\simeq\frac{U_0\sqrt{N}\alpha_0}{\bar{\Delta}_{c,j}+i\kappa}\int e^{\mp iq_jx}\hat{\psi}^\dagger(x)\hat{\psi}(x)dx+\mathcal{O}\left(\frac{1}{\Delta_a^2}\right),
\end{aligned}
\end{align}
where $\bar{\Delta}_{c,j}=\omega-\omega_j-NU_0$ and we neglect the $\mathcal{O}(1/\Delta_a^2)$ terms. Inserting expressions (\ref{eq:ao_op}) and (\ref{eq:aj_op}) into the Heisenberg equation (\ref{eq:heiseqs1}) for $\hat{\psi}(x)$, leads to
\begin{widetext}
\begin{align}\label{eq:Heiseq_atomonly}
\begin{aligned}
\frac{\partial \hat{\psi}(x)}{\partial t}&=\left[-\frac{1}{2m}\frac{\partial^2}{\partial x^2}+V_1(x)+\int\hat{\psi}^\dagger(x')V_2(x,x')\hat{\psi}(x')dx'\right]\hat{\psi}(x),
\end{aligned}
\end{align}
\end{widetext}
where the atom-atom interaction is given by:
\begin{align}\label{eq:twobodypot}
\begin{aligned}
V_2(x,x')&=\frac{\eta^2}{\bar{\Delta}_c^2+\kappa^2}\left\{ \frac{U_0}{N}+4U_0^2\sum_{q_j\in S}\frac{\bar{\Delta}_{c,j}}{\bar{\Delta}_{c,j}^2+\kappa^2}\cos[q_j(x-x')] \right\},
\end{aligned}
\end{align}
and the constant first term in (\ref{eq:twobodypot}) can be neglected. An effective atom-only Hamiltonian for this case is then given by Eq. (\ref{eq:eff_ham3_2}).

\section{Appendix C: Excitation matrix $\hat{\mbox{M}}$}
Inserting $\psi(x,t)=e^{-i\mu_0 t}[\psi_0(x)+\delta\psi(x,t)]$ and $\alpha_j(t)=\alpha_j^0+\delta\alpha_j(t)$, for $j=0,\pm$, into Eq. (\ref{eq:eqsallfull}), the dynamical equations are:
\begin{widetext}
\begin{align}\label{eq:eqsallperturb1}
\begin{aligned}
\frac{\partial \delta\alpha_0}{\partial t}   &=(i\bar{\Delta}_c-\kappa) \delta\alpha_0 -iu_0[\mathcal{I}_+^0\delta\alpha_++\mathcal{I}_-^0\delta\alpha_-+\int A_0(\psi_0\delta\psi^*+\psi_0^*\delta\psi)dx],\\
\frac{\partial \delta\alpha_\pm}{\partial t}  &=(i\bar{\Delta}_c'-\kappa) \delta\alpha_\pm -iu_0[\mathcal{I}_\mp^0\delta\alpha_0+\mathcal{I}_{2\mp}^0\delta\alpha_\mp+\int A_\pm(\psi_0\delta\psi^*+\psi_0^*\delta\psi)dx],\\
 i\frac{\partial \delta\psi}{\partial t}  &=(\mathcal{H}_0-\mu_0)\delta\psi+u_0\psi_0(A_0\delta\alpha_0^*+A_0^*\delta\alpha_0+A_+\delta\alpha_+^*+A_+^*\delta\alpha_++A_-\delta\alpha_-^*+A_-^*\delta\alpha_-),
 \end{aligned}
\end{align}
\end{widetext}
where $A_0=\alpha_+^0e^{iq_cx}+\alpha_-^0e^{-iq_cx}$, $A_\pm=\alpha_0^0e^{\mp iq_cx}+\alpha_\mp^0e^{\mp 2iq_cx}$, and the Hamiltonian $\mathcal{H}_0$ is given by: 
\begin{widetext}
\begin{align}\label{eq:h0}
\begin{aligned}
\mathcal{H}_0 &=-\frac{1}{2m}\frac{\partial^2}{\partial x^2}+ u_0\{[(\alpha_0^0)^*\alpha_+^0+(\alpha_-^0)^*\alpha_0^0]e^{iq_cx}
+[(\alpha_0^0)^*\alpha_-^0+(\alpha_+^0)^*\alpha_0^0]e^{-iq_cx}
+(\alpha_-^0)^*\alpha_+^0e^{2iq_cx}+(\alpha_+^0)^*\alpha_-^0e^{-2iq_cx}\}.
\end{aligned}
\end{align}
\end{widetext}
Writing $\delta\psi(x,t)=\delta\psi^+(x)e^{-i\Omega t}+[\delta\psi^{-}(x)]^*e^{i\Omega^* t}$ and $\delta\alpha_j(t)=\delta\alpha_j^+e^{-i\Omega t}+[\delta\alpha_j^{-}]^*e^{i\Omega^* t}$, and collecting the $e^{-i\Omega t}$, $e^{i\Omega^* t}$ terms, we get the equations:
\begin{widetext}
\begin{align}\label{eq:eqsallperturb_eigenfinal}
\begin{aligned}
\Omega\delta\alpha_0^+ &=(-\bar{\Delta}_c-i\kappa) \delta\alpha_0^+ +u_0[\mathcal{I}_+^0\delta\alpha_+^++\mathcal{I}_-^0\delta\alpha_-^++\int A_0(\psi_0\delta\psi^-+\psi_0^*\delta\psi^+)dx],\\
\Omega\delta\alpha_0^- &=(\bar{\Delta}_c-i\kappa) \delta\alpha_0^- -u_0[(\mathcal{I}_+^{0})^*\delta\alpha_+^-+(\mathcal{I}_-^0)^*\delta\alpha_-^-+\int A_0^*(\psi_0\delta\psi^-+\psi_0^*\delta\psi^+)dx],\\
\Omega \delta\alpha_+^+ &=(-\bar{\Delta}_c'-i\kappa) \delta\alpha_+^+ +u_0[\mathcal{I}_-^0\delta\alpha_0^++\mathcal{I}_{2-}^0\delta\alpha_-^++\int A_+(\psi_0\delta\psi^-+\psi_0^*\delta\psi^+)dx],\\
\Omega \delta\alpha_+^- &=(\bar{\Delta}_c'-i\kappa) \delta\alpha_+^- -u_0[(\mathcal{I}_-^0)^*\delta\alpha_0^-+(\mathcal{I}_{2-}^0)^*\delta\alpha_-^-+\int A_+^*(\psi_0\delta\psi^-+\psi_0^*\delta\psi^+)dx],\\
\Omega \delta\alpha_-^+ &=(-\bar{\Delta}_c'-i\kappa) \delta\alpha_-^+ +u_0[\mathcal{I}_+^0\delta\alpha_0^++\mathcal{I}_{2+}^0\delta\alpha_+^++\int A_-(\psi_0\delta\psi^-+\psi_0^*\delta\psi^+)dx],\\
\Omega \delta\alpha_-^- &=(\bar{\Delta}_c'-i\kappa) \delta\alpha_-^--u_0[(\mathcal{I}_+^0)^*\delta\alpha_0^-+(\mathcal{I}_{2+}^0)^*\delta\alpha_+^-+\int A_-^*(\psi_0\delta\psi^-+\psi_0^*\delta\psi^+)dx],\\
 \Omega \delta\psi^+ &=(\mathcal{H}_0-\mu_0)\delta\psi^++u_0\psi_0(A_0\delta\alpha_0^-+A_0^*\delta\alpha_0^++A_+\delta\alpha_+^-+A_+^*\delta\alpha_+^++A_-\delta\alpha_-^-+A_-^*\delta\alpha_-^+),\\
 \Omega \delta\psi^- &=-(\mathcal{H}_0-\mu_0)\delta\psi^--u_0\psi_0^*(A_0\delta\alpha_0^-+A_0^*\delta\alpha_0^++A_+\delta\alpha_+^-+A_+^*\delta\alpha_+^++A_-\delta\alpha_-^-+A_-^*\delta\alpha_-^+).
 \end{aligned}
\end{align}
\end{widetext}
The Eq. (\ref{eq:eqsallperturb_eigenfinal}) is an eigenvalue problem that can readily be written in matrix form $\hat{\mbox{M}}\mathbf{R} =\Omega\mathbf{R}$, where $\mathbf{R}$ is given by Eq. (\ref{eq:perturbvector}).

\bibliography{references}
\clearpage
\newpage

\end{document}